\documentclass[aps,prl,twocolumn,showpacs,superscriptaddress,letterpaper]{revtex4}  
\usepackage{graphicx}  
\usepackage{dcolumn}   
\usepackage{bm}        
\usepackage{amsmath,latexsym,amssymb,amsthm,revsymb,array}
\usepackage{wrapfig,color}
\usepackage{hyperref}

\newcommand{\mm}{\,\mbox{mm}}

\newcommand{\nm}{\,\mbox{nm}}

\newcommand{\kHz}{\,\mbox{kHz}}

\newcommand{\mW}{\,\mbox{mW}}

\newcommand{\nc}{\newcommand}
\nc{\SR}{\mathrm{SR}}
\nc{\SE}{\mathrm{SE}}
\nc{\NS}{\mathrm{NS}}

\nc{\pw}{\mathrm{PW}}
\nc{\arbclass}{\mathrm{\Omega}}
\nc{\rnc}{\renewcommand}
\nc{\beq}{\begin{equation}}
\nc{\mc}{\mathcal}
\nc{\eeq}{{\end{equation}}}
\nc{\beqa}{\begin{eqnarray}}
\nc{\eeqa}{\end{eqnarray}}
\nc{\lbar}[1]{\overline{#1}}
\nc{\bra}[1]{\langle#1|}
\nc{\ket}[1]{|#1\rangle}
\nc{\ketbra}[2]{|#1\rangle\!\langle#2|}
\nc{\braket}[2]{\langle#1|#2\rangle}
\nc{\proj}[1]{| #1\rangle\!\langle #1 |}
\nc{\avg}[1]{\langle#1\rangle}
\nc{\Rank}{\operatorname{rank}\,}
\nc{\smfrac}[2]{\mbox{$\frac{#1}{#2}$}}
\nc{\tr}{\operatorname{Tr}}
\nc{\ox}{\otimes}
\nc{\catchset}{T}
\nc{\dg}{\dagger}
\nc{\dn}{\downarrow}
\nc{\cA}{{\cal A}}
\nc{\cB}{{\cal B}}
\nc{\cC}{{\cal C}}
\nc{\cD}{{\cal D}}
\nc{\cE}{{\cal E}}
\nc{\cF}{{\cal F}}
\nc{\cG}{{\cal G}}
\nc{\cH}{{\cal H}}
\nc{\cI}{{\cal I}}
\nc{\cJ}{{\cal J}}
\nc{\cK}{{\cal K}}
\nc{\cL}{{\cal L}}
\nc{\cM}{{\cal M}}
\nc{\cN}{{\cal N}}
\nc{\cO}{{\cal O}}
\nc{\cP}{{\cal P}}
\nc{\cR}{{\cal R}}
\nc{\cS}{{\cal S}}
\nc{\cT}{{\cal T}}
\nc{\cX}{{\cal X}}
\nc{\cY}{{\cal Y}}
\nc{\cZ}{{\cal Z}}
\nc{\csupp}{{\operatorname{csupp}}}
\nc{\qsupp}{{\operatorname{qsupp}}}
\nc{\var}{{\operatorname{var}}}
\nc{\rar}{\rightarrow}
\nc{\lrar}{\longrightarrow}
\nc{\polylog}{{\operatorname{polylog}}}
\nc{\1}{{\mathbbm{1}}}
\nc{\wt}{{\operatorname{wt}}}

\nc{\RR}{{{\mathbb R}}}
\nc{\CC}{{{\mathbb C}}}
\nc{\FF}{{{\mathbb F}}}
\nc{\NN}{{{\mathbb N}}}
\nc{\ZZ}{{{\mathbb Z}}}
\nc{\PP}{{{\mathbb P}}}
\nc{\QQ}{{{\mathbb Q}}}
\nc{\UU}{{{\mathbb U}}}
\nc{\EE}{{{\mathbb E}}}
\nc{\id}{{\operatorname{id}}}

\nc{\CHSH}{{\operatorname{CHSH}}}

\nc{\be}{\begin{equation}}
\nc{\ee}{{\end{equation}}}
\nc{\bea}{\begin{eqnarray}}
\nc{\eea}{\end{eqnarray}}
\nc{\<}{\langle}
\rnc{\>}{\rangle}
\nc{\Hom}[2]{\mbox{Hom}(\CC^{#1},\CC^{#2})}
\nc{\rU}{\mbox{U}}

\nc{\ob}[1]{#1}
\def\chn{\mc{N}_{\boxtimes}}

\def\tP{\mathrm{P}}

\def\SE{\mathrm{SE}}

\begin{document}

	\title{Entanglement-Enhanced Classical Communication over a Noisy Classical Channel}

     \author{R. Prevedel}\email{robert.prevedel@iqc.ca}

     \author{Y. Lu}

     \author{W. Matthews}
    
     \author{R. Kaltenbaek}
     
     \author{K.J. Resch}\email{kresch@iqc.ca}
     
     \affiliation{Institute for Quantum Computing, University of Waterloo, Waterloo, N2L 3G1, ON, Canada}	
	
	\date{\today}

	\begin{abstract}
	We present and experimentally demonstrate a communication protocol that employs shared entanglement to reduce errors when sending a bit over a particular noisy classical channel. Specifically, it is shown that, given a single use of this channel, one can transmit a bit with higher success probability when sender and receiver share entanglement compared to the best possible strategy when they do not. The experiment is realized using polarization-entangled photon pairs, whose quantum correlations play a critical role in both the encoding and decoding of the classical message. Experimentally, we find that a bit can be successfully transmitted with probability $0.891\pm0.002$, which is close to the theoretical maximum of $(2+2^{-1/2})/3\approx0.902$ and is significantly above the optimal classical strategy, which yields $5/6\approx0.833$. 	
	\end{abstract}

	\maketitle

Two parties that share an entangled quantum system can achieve communication tasks which would otherwise be impossible: Sending two bits of classical information using only one qubit~\cite{Bennett92b}, unconditionally secure communication~\cite{Bennett1984}, transferring quantum information from one quantum system to another ~\cite{Bennett1993,Bouwmeester97}, and reducing communication complexity in distributed computations~\cite{cc}.

It is much less studied how entanglement can assist in sending classical information over a classical channel. The main result in this context is that entanglement cannot increase the capacity of any classical channel~\cite{Bennett1993,Bennett1999} in the sense of Shannon~\cite{Shannon1948}. However, it has recently been shown that the number of different messages that can be sent error-free with a \emph{single} use of a \emph{noisy} classical channel (i.e. the \emph{one-shot zero-error capacity}) can be increased when the sender and receiver share entanglement \cite{Cubitt2010}. Another interpretation of this result is that for a fixed number of possible messages and a fixed number of channel uses, entanglement can be used to increase the probability of successful decoding, which raises interesting general questions about when entanglement can assist in this way.

In this Letter we describe an example of such an entanglement-enhanced classical communication protocol that, compared to the example given in \cite{Cubitt2010}, exhibits a larger absolute gap in the assisted and unassisted success probabilities, involves a much simpler classical channel, a smaller entangled state and, most significantly, is experimentally feasible. We go on to describe an experimental implementation which clearly demonstrates the advantage gained from using entanglement experimentally. This represents an application of entanglement in a new setting, where it is surprising that one can benefit from quantum effects.

We first briefly discuss error-correcting codes for classical channels with and without entanglement assistance. Then, for a specific classical channel $\chn$, we show that the maximum success probability for sending one bit with a single use of this channel is $5/6\approx0.833$, even if the parties have shared randomness at their disposal. We then show that if Alice and Bob share a pair of maximally entangled qubits, the sucess probability can be increased to $(2+2^{-1/2})/3\approx0.902$. Using entangled photons we then implement this protocol experimentally, achieving a success probability of $0.891\pm0.002$. This is close to the theoretical value and significantly better than the best classical strategy.

\emph{Classical vs. entanglement-assisted codes.}
The task we are studying is this: Alice wants to be able to send one of $M$ possible messages to Bob, by making one use of a noisy classical channel $\mc{N}$ (and no other signals). The channel has a finite set of input symbols, and a finite set of output symbols, and is described by the condition probabilities of the output symbols given the input symbol.  Given a uniform prior distribution on the messages, we want to maximise the probability that Bob determines the message correctly.

If we have a purely classical protocol which depends on some random variable $R$, its success probability will be the mean of the success probabilities of the protocols for fixed values of $R$, and therefore no larger than the best success probability attained by some fixed value of $R$ and we can eliminate the dependence on $R$ without detriment. Therefore, there is an optimal classical protocol which is deterministic: Alice maps each messages to some input symbol, and Bob has a decoding rule that maps output symbols to messages (if the channel is of the form $\mc{M}^{\ox k}$, corresponding to $k$ independent uses of a given channel $\mc{M}$, then such a protocol is called a \emph{block code} of \emph{block length} $k$ for the channel $\mc{M}$). This is true even if $R$ is \emph{shared} randomness i.e. its value is known to both parties.

\begin{figure}[t]
	\includegraphics[scale=0.5]{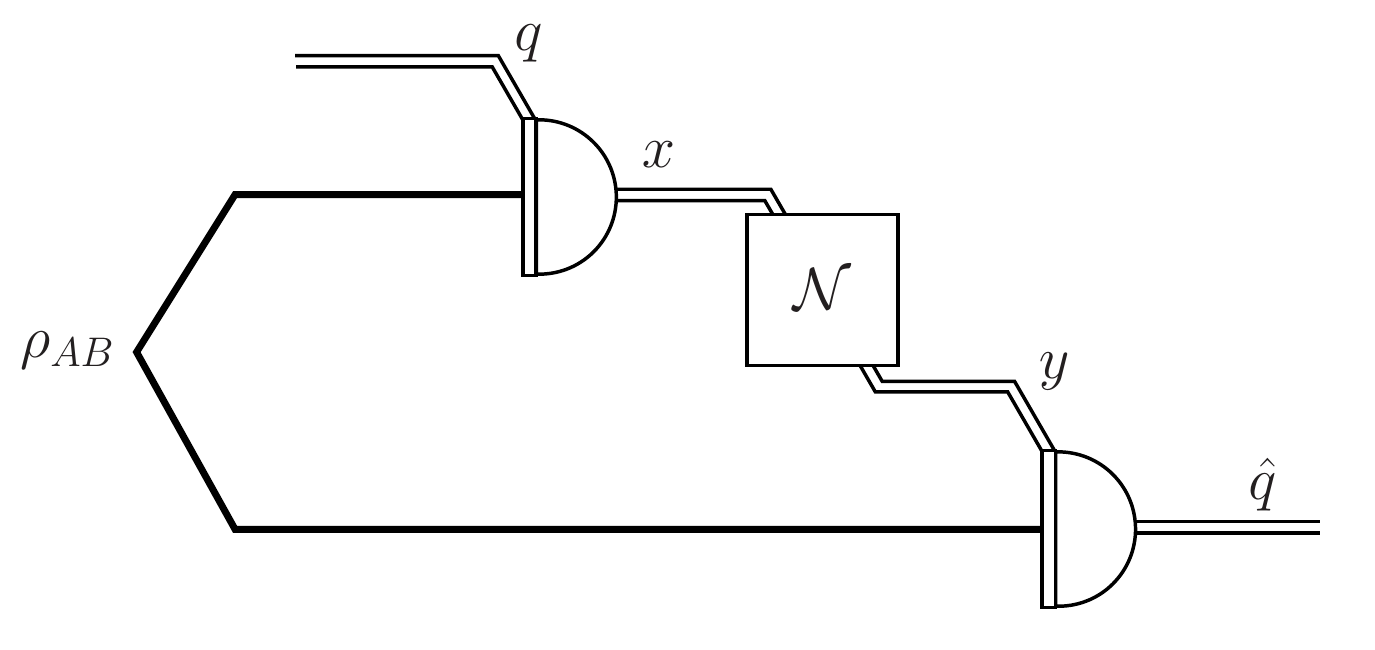}
	\caption{General form of an entanglement-assisted coding protocol for sending classical message $q$ with a single use of a classical channel $\mc{N}$ and an entangled state $\rho_{AB}$. Double lines represent classical communication.}
	\label{EAC}
\end{figure}

On the other hand, we will see that shared \emph{entanglement} can increase the probability of success beyond any classical protocol. The most general form of an entanglement-assisted protocol to send one of $M$ messages using a classical channel is illustrated in Fig~\ref{EAC}: Alice and Bob possess systems $A$ and $B$ in some entangled state $\rho_{AB}$ (note that a separable state can be simulated by shared randomness). The message $q$ determines which measurement Alice performs on her system $A$ and the measurement outcome determines the input $x$ she makes to the channel. This does not imply that the channel input can not also depend on $q$, but this dependence can be absorbed into the definition of the measurement. The channel output $y$ then determines the measurement Bob does on his system ($B$) and its outcome determines his decoding $\hat{q}$ of the message (again, the definition of the measurement can incorporate any dependence on $y$). For more details see \cite{Cubitt2010} and \cite{Cubitt2010b}.


\begin{figure}[t]
	\includegraphics[scale=0.45]{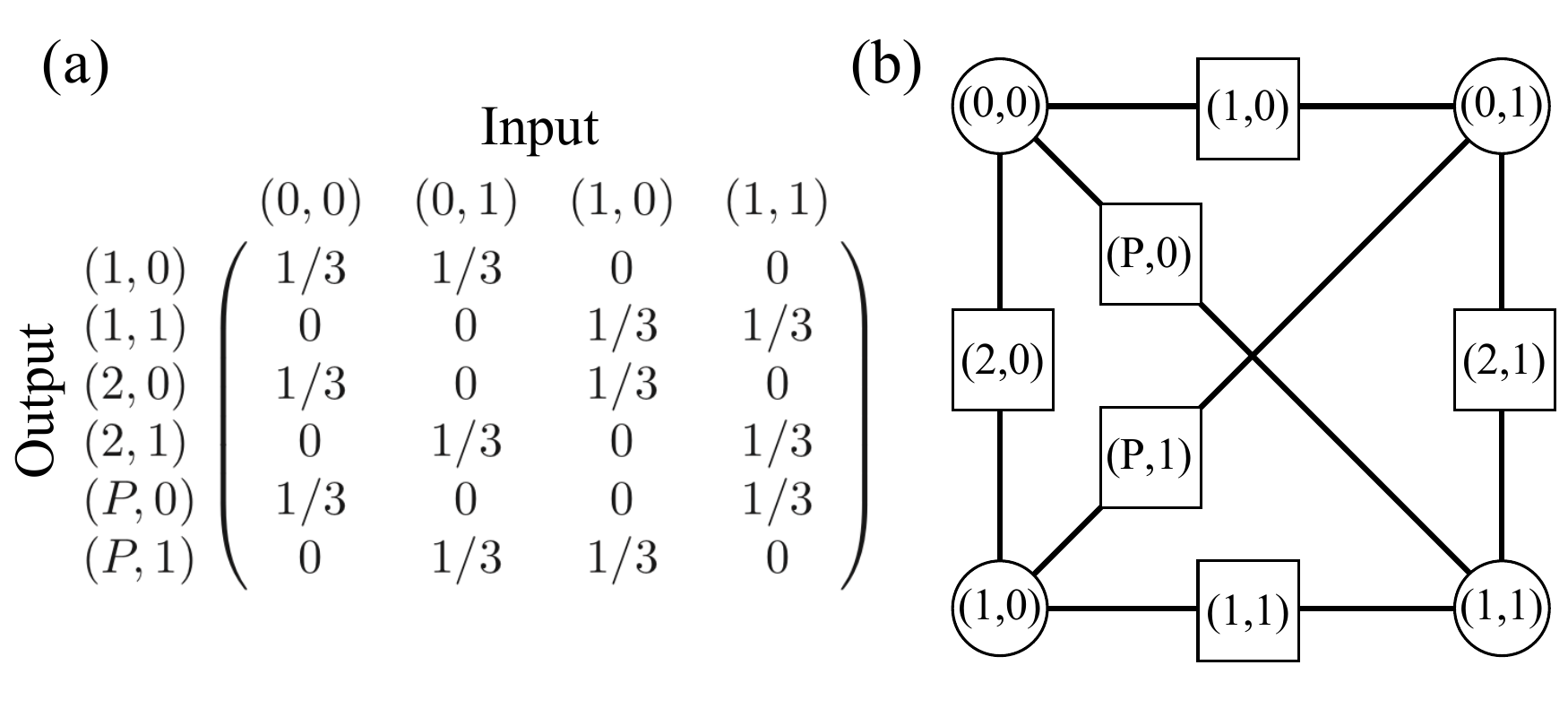}
	\caption{The conditional probability matrix of $\chn$ (a) and a graph representing its structure (b). The inputs (outputs) are labeled as circles (squares).}
	\label{chan}
\end{figure}

\emph{The channel.}
The particular noisy classical channel $\chn$ we consider in this paper takes a two-bit input, $(b_1,b_2)$. It produces an output $(t,b)$ consisting of a trit $t$ which is selected at random from the set $\{ 1,2,P \}$ (each with probability $1/3$), and a bit $b$ which is equal to $x_1$ when $t = 1$, $x_2$ when $t = 2$, and is the parity $x_1 \oplus x_2$ when $t = P$. The conditional probability matrix $\Pr\left[(t,b)|(b_1,b_2)\right]$ of $\chn$ is shown in Figure \ref{chan}a. In Figure \ref{chan}b, the four input symbols are represented as the circular vertices of a graph whose edges (with square labels) correspond to the 6 output symbols. In terms of this picture, when an input is made, the output is chosen uniformly at random from the three edges in the graph which are incident with it.

A classical code to send a bit with one use of $\chn$ uses two of the four possible inputs to represent the two messages `$0$' and `$1$'. For example we might use $(0,1)$ to encode $0$ and $(1,0)$ for $1$. The optimal decoding map is easy to see: If the output is $(1,0)$ or $(2,1)$ then the input was certainly $(0,1)$ so Bob decodes `$0$'. Likewise, he decodes $(1,1)$ and $(2,0)$ to message `$1$', and this is also always correct. Output $(P,0)$ never occurs. If the output is $(P,1)$ then the two messages are equally probable, so the best Bob can do is guess. He will be right with probability 1/2, and this output occurs with probability $1/3$ so he is wrong with probability $1/6$. The symmetry of the channel (which is evident in Figure \ref{chan}b) means that any other assignment is equivalent to this one under relabelling of the channel outputs, and has the same probability of successful decoding, namely $5/6$.

\emph{The entanglement-assisted protocol.}
Shared entanglement allows to improve on the optimal classical strategy described in the last section. The protocol assumes that the two parties each have one subsystem from a maximally polarization-entangled state of two photons $\ket{\Phi^+} = (\ket{H H} + \ket{V V})/\sqrt{2}$. For the protocol to proceed, each party performs a measurement in one out of two bases on their photon. Each measurement has two possible outcomes, projecting on quantum states $\ket{\theta} = \cos\theta \ket{H} + \sin\theta \ket{V}$. We will represent each such measurement by $(\ket{\theta_0}, \ket{\theta_1})$ to denote the (orthogonal) outcomes $0$ and $1$ for each measurement, respectively.

For Alice, measurement $0$ is $(\ket{\pi/4},\ket{3\pi/4})$ and measurement $1$ is $(\ket{0},\ket{\pi/2})$. Bob's measurement $0$ is $(\ket{\pi/8},\ket{5\pi/8})$ and measurement $1$ is $(\ket{3\pi/8},\ket{7\pi/8})$. Note that these are the measurement settings that give rise to a maximal violation of a CHSH-inequality~\cite{Clauser69}. Alice performs measurement $q=0$ or $1$ and obtains outcome $\alpha$ while Bob makes measurement $v=0$ or $1$ and obtains outcome $\beta$. Here $\alpha$ and $\beta$ are bit values corresponding to the measurement outcomes. The marginal distributions of $\alpha$ and $\beta$ are uniform, independent of the inputs and the correlations will be such that the relation \cite{Popescu1994}
\begin{equation}\label{relation}
	\alpha \oplus \beta = qv
\end{equation}
holds with probability $\omega = (1+2^{-1/2})/2$ \footnote{The `Popescu-Rohrlich box' \cite{Popescu1994} is a notional bipartite non-signalling correlation which \emph{always} satisfies this relation. If a Popescu-Rohrlich box could be used in place of measurements on an entangled state, then the bit would be transmitted with certainty (see \cite{Cubitt2010b}).}. To send the bit $q$, Alice performs measurement $q$, and records the outcome $\alpha$. She then uses $(q,\alpha)$ as the input to $\chn$. In Table \ref{bobtab} and its caption we describe Bob's part of the protocol and how the success probability of $(2+2^{-1/2})/3 \approx 0.902$ can be achieved\footnote{It is interesting to note that part of the channel input, the bit $\alpha$, is random and independent of the message.}.

\begin{table}[b!]
\begin{tabular}{| c | c | c | c | c || c | c |}
	
\hline
	$t$	&	$b$				&	Bob chooses $v =$	&	 $\beta$			 &	 $\hat{q}$ 				 & $X_{\text{on}}$ & $Z_{\text{on}}$\\
\hline
	$1$	&	$q$				&	irrelevant		&	n/a		 &					 $b$							 &	 n/a	&	n/a	\\
	$2$	&	$\alpha$			&					1		 &	 $q\oplus \alpha$	 &	 $b\oplus\beta$		&	$1 \oplus b$	&	 $b$	\\
	$\tP$	&	$q\oplus\alpha$	&					0		 &	 $\alpha$			 &	 $b\oplus\beta$		&	$b$	 &	$b$	\\
\hline
\end{tabular}
\caption{The rows of this table track the protocol from left to right for the 3 possible values of $t$: Bob obtains $(t,b)$ from the channel, $b$'s dependence on the message $q$ and Alice's measurement outcome $\alpha$ being fixed by $t$ as described in the channel section; Based on $t$, Bob determined his measurement choice $v$. He gets the outcome $\beta$ - in this column we assume that the relation (\ref{relation}) \emph{does} hold. Substituting the choice Bob makes for $v$ in this relation determines $\beta$ as a function of $\alpha$ and $q$. The next column shows the function of $b$ and $\beta$ Bob calculates to get his decoding of the message $\hat{q}$. When the relation (\ref{relation}) holds, this function is equal to $q$ for all $t$. If the relation does not hold, then if $t = 1$ (with probability $1/3$) he is still correct. Therefore the probability of successful decoding is $\omega + (1-\omega)/3 = (2+2^{-1/2})/3$. The last two columns show the settings of the Pockels cells used to implement Bob's measurement in the experiment (1 is `ON', 0 is `OFF').}\label{bobtab}
\end{table}

\begin{figure}[t]
\includegraphics[width=\columnwidth]{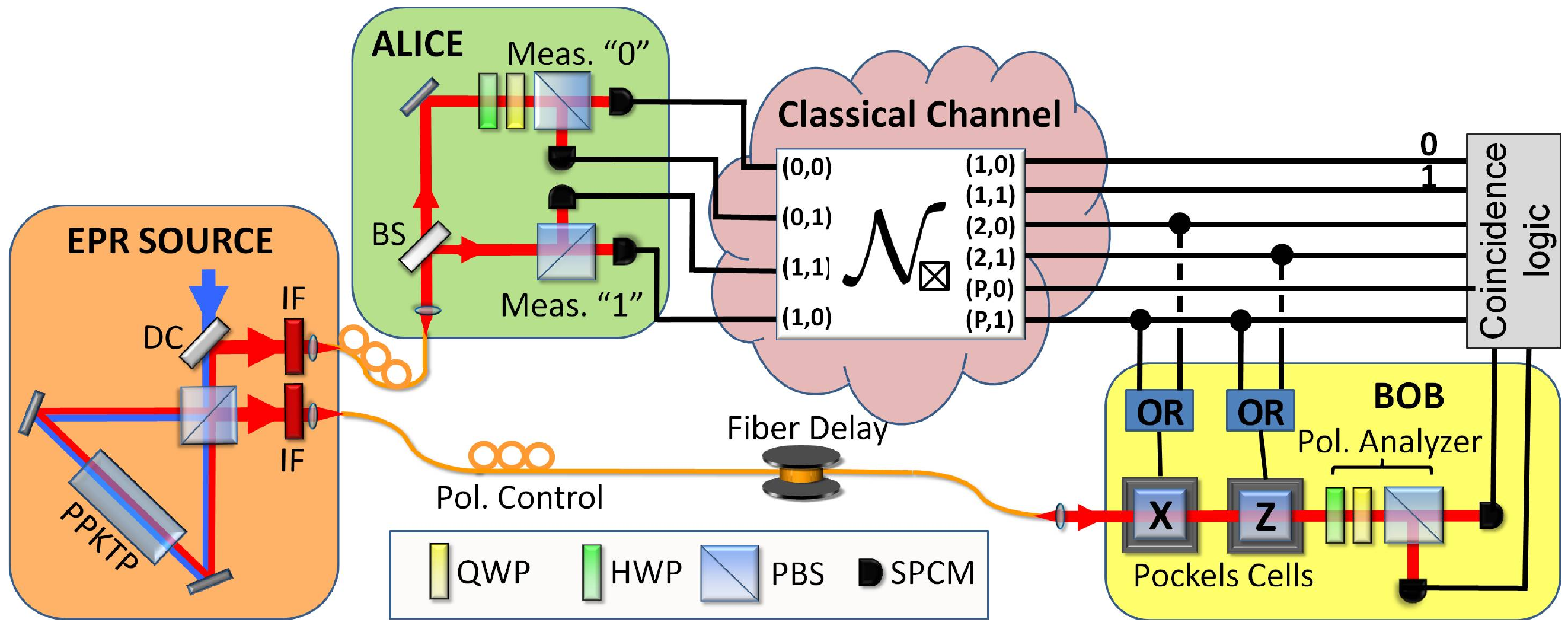}
   \caption{Experimental implementation. The Sagnac-SPDC source generates maximally entangled photon pairs of the form $\ket{\Phi^+}$, which are coupled into single-mode fibers (SMF). One of the photons is brought to Alice's side, where two polarization analyzer modules ($\ket{H/V}$ and $\ket{+/-}$ basis respectively) are separated by a 50/50 beamsplitter (BS). The classical channel $\chn$ performs its mapping dependent on the RNG operated at $200\kHz$. This ensures that every new input to the channel will likely encounter a different random signal. Depending on the output of $\chn$, fast Pockels cells perform $\sigma_x$ (X) or $\sigma_z$ (Z) on Bob's photon, which is delayed in a 50 m SMF to account for the feed-forward time.} \label{fig:setup}
\end{figure}

\emph{Experimental realization.}
The setup of our experiment is shown in Fig.~\ref{fig:setup}. We implement the classical channel, $\chn$, with its four inputs and six outputs by using an electronic circuit (CPLD logic chip Xilinx Xc2c64a) that is controlled by a random trit. This random trit, $t$, is generated at $200\kHz$ by a computer using National Instruments' LabView's pseudo-random number generator (RNG). We generate the entangled resource state for our communication protocol by using type-II spontaneous parametric down-conversion (SPDC). A 0.7$\mW$ diode laser at 404$\nm$ pumps a 25$\mm$  periodically-poled KTiOPO$_4$ (PPKTP) crystal in a Sagnac configuration, emitting entangled photons which are subsequently single-mode fiber-coupled after 3$\nm$ bandpass interference filters (IF) \cite{Wong2006,Biggerstaff2009}. Typically we observe a coincidence rate of 15 $\kHz$ directly at the source.

Alice's message, $q$, is selected randomly by means of a beamsplitter; if her photon is transmitted she performs measurement 0, and if the photon is reflected she performs measurement 1. Which of the four detectors fires determines the input to $\chn$ as described in the last section and illustrated in Fig.~\ref{fig:setup}. Depending on the output $y$ of the channel, Bob needs to actively choose his measurement basis and may need to invert his measurement outcome in order to decode the message. Both of these actions are implemented using two fast RbTiOPO$_4$ (RTP) Pockels cells (PC), aligned so as to perform a $\sigma_x$ (X) and $\sigma_z$ (Z) operation, respectively~\cite{Prevedel2007,Biggerstaff2009}. The states $X_{\text{on}}$ and $Z_{\text{on}}$ of the Pockels cells, which can each independently be $0$ or $1$, are shown in Table \ref{bobtab}.

After passing the Pockels cells, Bob's photon is measured in the $(\ket{\pi/8},\ket{5\pi/8})$ basis, where the output $\beta$ is $0$ or $1$ depending on whether the photon is detected in the transmitted or reflected output port of the analyzer module, respectively. Unless $t=1$, Bob uses the outcome of the measurement in combination with the output $b$ of the channel to decode the message. If $t=1$, Bob ignores the measurement result and directly uses $b$ as the decoded message $\hat{q}$ (see Table \ref{bobtab}). The measurement bases of Alice and Bob are set by half- and quarter-wave plates (HWPs, QWPs) followed by polarizing beam splitters (PBSs).

\begin{figure}[t]
\includegraphics[width=\columnwidth]{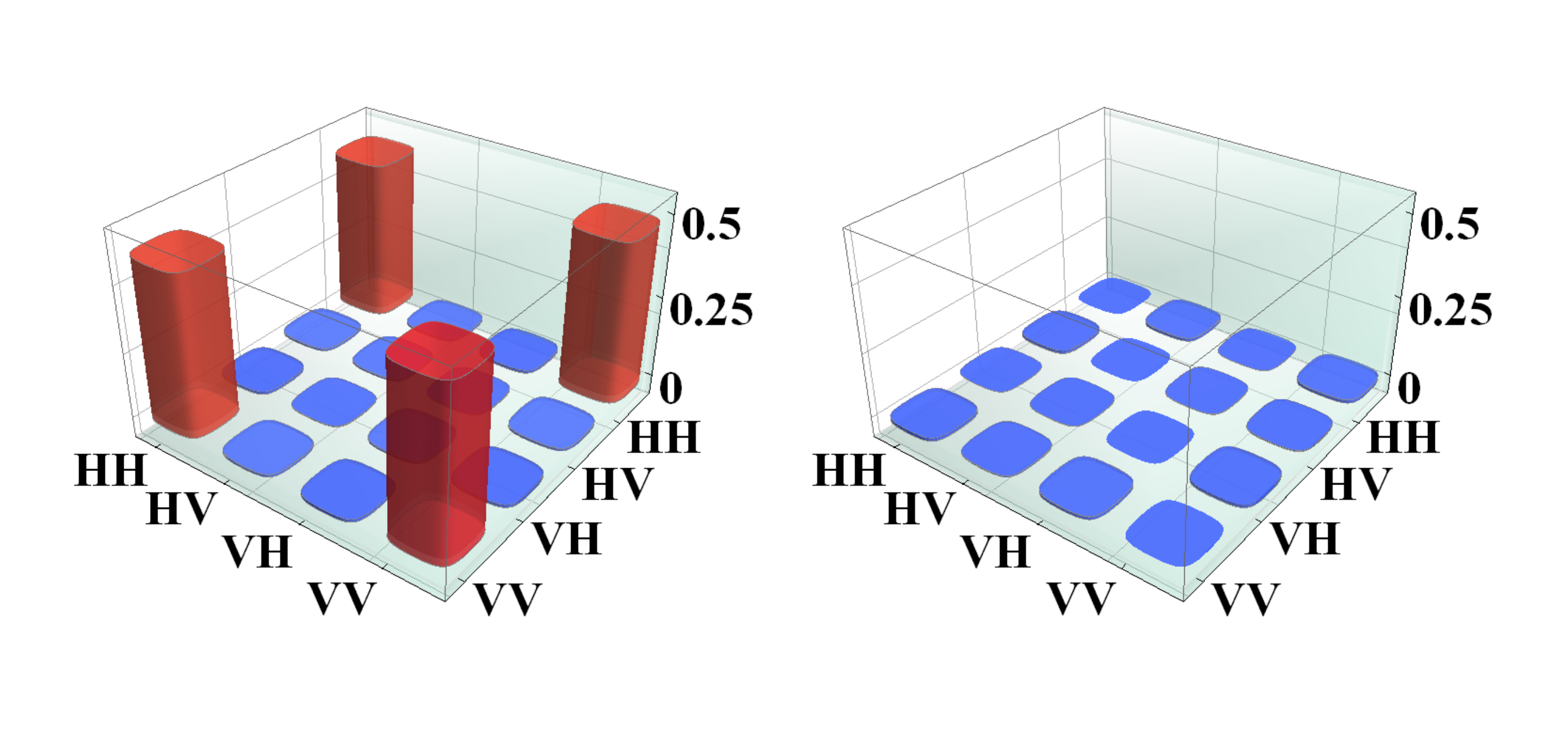}\\
\vspace{-1mm}
\includegraphics[width=\columnwidth]{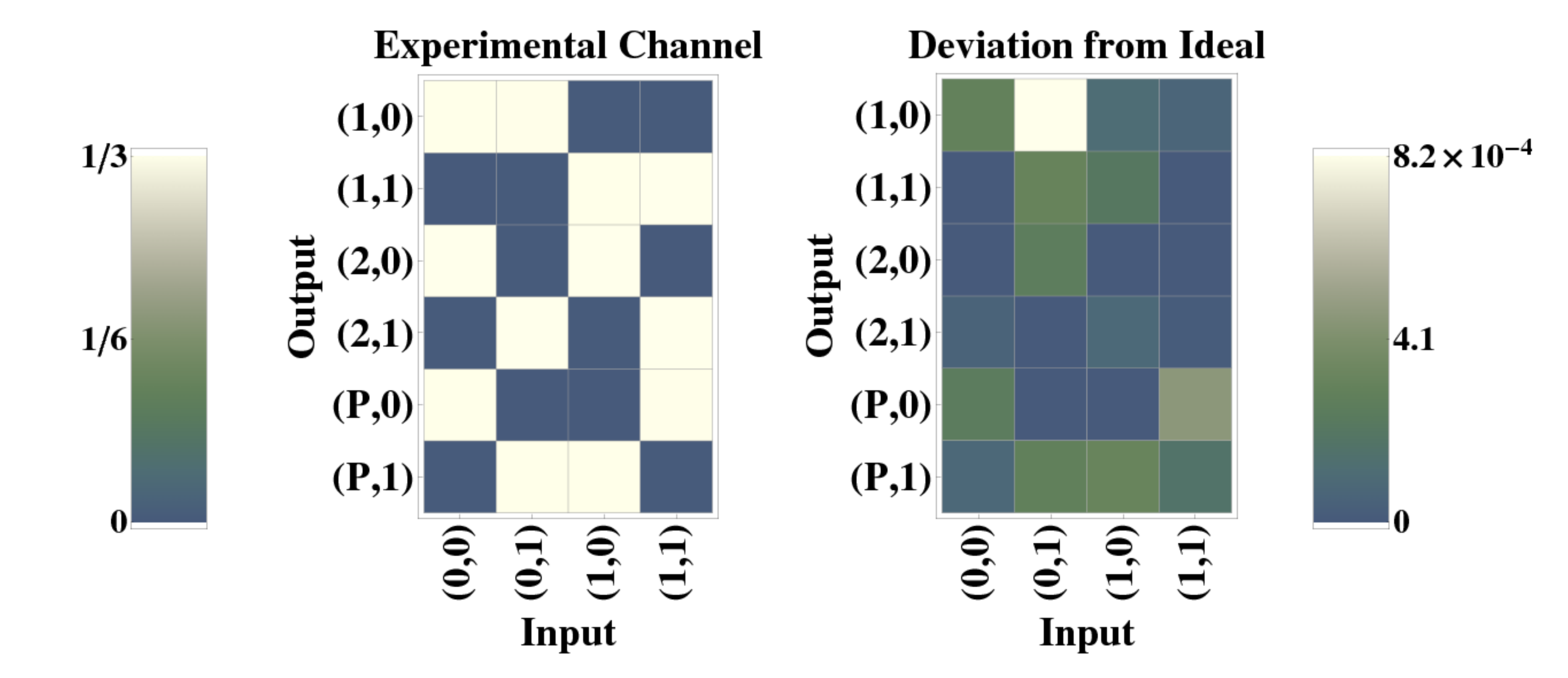}
   \caption{Source and channel characterization. Top: Experimentally reconstructed density matrix of our two-photon entangled state $\rho_{\text{exp}}$: real part (left panel) and imaginary part (right panel). It has fidelity $F=0.981\pm 0.001$ with the ideal $\ket{\Phi^+}$ and a tangle of $\tau=0.925\pm0.004$. The error bars for these results were calculated by a 200 run Monte-Carlo simulation, adding Poissonian noise to the count statistics in each run. Bottom: Truth tables for the experimental channel (left panel) and its deviation from an ideal one (right panel - c.f. Fig. \ref{chan}(a)), as recorded during our experiment.}
   \label{fig:state}
\end{figure}

To evaluate the quality of our entangled state, we performed quantum state tomography~\cite{James2001,Langford2005}. On Alice's side, we analyzed the photons in the transmitted arm of the BS. We recorded coincidences between this output and Bob's polarization analyzer following the (switched off) PCs. Coincidence measurements were integrated over 8 s for each of 36 different measurements, comprising all combinations of the six eigenstates of $X$, $Y$, and $Z$ on Alice's and Bob's qubit, respectively. Using a maximum-likelihood technique~\cite{James2001} we reconstructed the density matrix shown in Fig.~\ref{fig:state}.

To characterize the performance of the channel during the experiment, we record the coincidence events for all possible input/output combinations. The frequencies of the outputs for each of the inputs are shown in a ``truth table'' in Fig.~\ref{fig:state}. We can quantify the overlap between the measured truth table, $N_{\text{exp}}$ and the ideal truth table, $N_{\text{th}}$, using the so-called inquisition~\cite{White2007}, $I=\text{Tr}(N_{\text{exp}}N_{\text{th}}^T)/\text{Tr}\left(N_{\text{th}}N_{\text{th}}^T\right)$. For our channel, $I=0.9992\pm 0.0001$, where the uncertainty was calculated by a Monte-Carlo simulation with binomially distributed random-signal frequencies (which stem from the RNG in the experiment) added in each run.

\begin{figure}[t!]
\includegraphics[width=7.2cm]{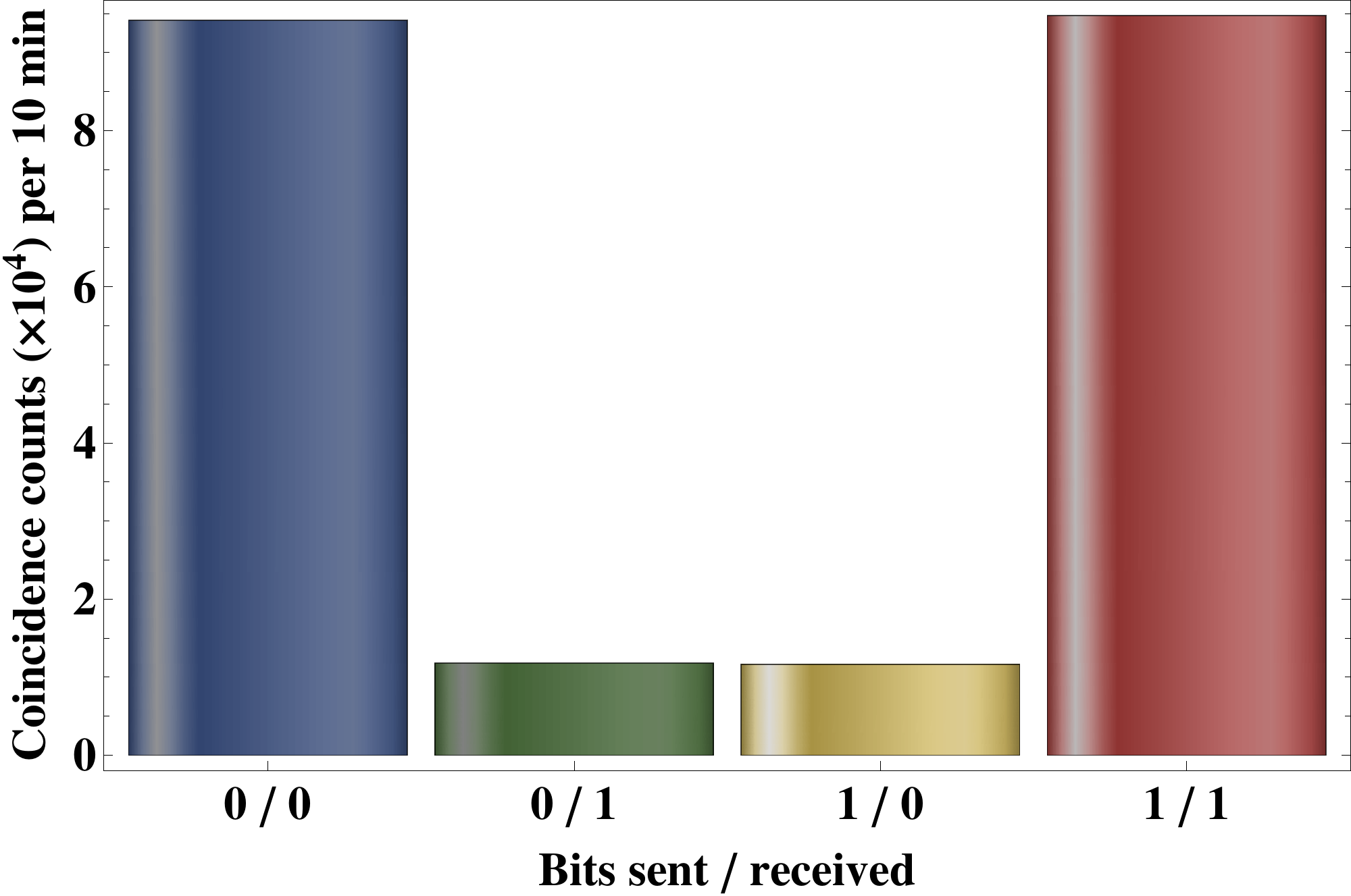}
   \caption{Experimental results. The bar chart displays coincidences recorded (input bit/decoded bit) over a period of 10 minutes. In total, we recorded $188845$ ($23429$) bits that were successfully (unsuccessfully) transmitted. This data yields a success probability $P_{\text{exp}}=0.891\pm 0.002$.} 
   \label{fig:result}
\end{figure}

\emph{Experimental results.}
In our experiment, we record all combinations of coincidence counts between Alice's four single-photon counting detectors and the two on Bob's side. This allows us to obtain the success probability $P_{\text{exp}}$, i.e. the ratio of successfully received and decoded bits over the total number of bits sent. The results are shown in Fig.~\ref{fig:result}. From the counts recorded over 10 min we calculate $P_{\text{exp}}=0.891\pm 0.002$, where the error bar stems from Poissonian count statistics. Imperfect state creation and feed-forward operations lead to a decrease from the ideal, theoretical success probability, $P_{\text{th}}=(2+2^{-1/2})/3\approx0.902$. The tomographic data of our entangled resource state allows us to calculate the \emph{expected} success probability of our protocol, which we infer as $0.8957\pm 0.0004$. This shows that our implementation of the protocol is mostly limited by the quality of our entangled resource state and that the rest of the setup is operating at high fidelity.

\emph{Conclusion.}
In summary, we have shown how shared entanglement can be used to improve the performance of a completely classical communication task, namely sending one bit with a single use of a noisy classical channel. This scheme shows how entanglement can offer a distinct advantage in a classical error coding scenario involving a finite number of channel uses.  Our results lead to interesting questions for further study: which classical communication channels can benefit from entanglement and by how much? For instance, can we find general bounds on the gap between the error probability with and without entanglement assistance? How do these ideas generalize in the context of multi-terminal communication assisted by multi-partite entanglement?

We thank D. Hamel, J. Lavoie, M. Piani and R. Spekkens for valuable discussions, and Z. Wang for building the channel logic. We are grateful for financial support from Ontario Ministry of Research and Innovation ERA, QuantumWorks, NSERC, OCE, Industry Canada and CFI. R.P. acknowledges support by MRI and the Austrian Science Fund (FWF).

\bibliographystyle{natbib}

\end{document}